\documentclass[a4paper,fleqn]{cas-sc}

\usepackage[authoryear]{natbib}
\usepackage{url}
\usepackage{hyperref}

\usepackage{caption}
\usepackage{subcaption}

\def\tsc#1{\csdef{#1}{\textsc{\lowercase{#1}}\xspace}}
\tsc{WGM}
\tsc{QE}
\tsc{EP}
\tsc{PMS}
\tsc{BEC}
\tsc{DE}

\begin{document}
\let\WriteBookmarks\relax
\def\floatpagepagefraction{1}
\def\textpagefraction{.001}

\shorttitle{Impact of Cognitive Dissonance on Social Hysteresis}

\shortauthors{B Kami\'nska et~al.}

\title [mode = title]{Impact of Cognitive Dissonance on Social Hysteresis: Insights from the Expressed and Private Opinions Model.} 
\tnotemark[1]

\tnotetext[1]{This document is the results of the research project funded by the National Science Centre
Poland Grant 2019/35/B/HS6/02530.}

\author[1]{Barbara Kami\'{n}ska}[orcid=0000-0003-0663-8753]

\fnmark[1]

\ead{b.kaminska@pwr.edu.pl}
\credit{Methodology, Software}

\author[2]{Katarzyna Sznajd-Weron}[orcid=0000-0002-1851-8508]

\cormark[2]
\ead{katarzyna.weron@pwr.edu.pl}

\credit{Conceptualization of this study,Writing - Original draft preparation}
\cortext[cor1]{Corresponding author}
\cortext[cor2]{Principal corresponding author}

\affiliation[1,2]{organization={Wroclaw University of Science and Technology, Faculty of Management},
    addressline={Wyb. Wyspiańskiego 27}, 
    city={Wroclaw},
    postcode={50-370}, 
    country={Poland}}

\begin{abstract}
The growing interest in models of opinion dynamics, particularly those that distinguish between private beliefs and publicly expressed opinions, spans several academic disciplines, from the social sciences to the hard sciences. They have been developed to study decision-making mechanisms and applied to many social phenomena, such as pluralistic ignorance, the spiral of silence, or preference falsification. Within these models, however, there is a notable gap in the study of social hysteresis, a concept crucial to understanding the delayed responses of societies to the rapid changes of the modern world. This research aims to fill this gap by examining the impact of cognitive dissonance on social hysteresis through an updated model of expressed and private opinions (EPOs). We study the model both analytically and through Monte Carlo simulations. To facilitate understanding and replication of our model, and to provide a resource for further exploration to a diverse interdisciplinary audience, we have made the special NetLogo implementation of the model publicly available at \href{https://barbarakaminska.github.io/q-voter-model-expressed-private/}{GitHub}. By incorporating a cognitive dissonance mitigation mechanism into an agent-based $q$-voter-type model of Expressed and Private Opinions (EPOs), we demonstrate that this mechanism induces social hysteresis. Consequently, we argue that refraining from rationalizing publicly expressed opinions could mitigate social hysteresis and facilitate consensus, offering insights into potential strategies for managing societal responses to change.
\end{abstract}


\maketitle

\section{Introduction}
People naturally try to minimize cognitive dissonance, which occurs when their beliefs or publicly expressed opinions conflict with their actions \citep{fes:57}. However, this doesn't always lead to a change in behavior. Instead, they often rationalize their actions to reduce the cognitive dissonance. For example, individuals who are environmentally conscious at home may engage in vacation behaviors that harm the environment. Studies have found that while this attitude-behavior gap makes them uncomfortable, it doesn't necessarily cause them to change their behavior; instead, they offer justifications for their tourist activities \citep{Juv:Dol:14}. Another example is eating habits, where meat eaters may feel conflicted when confronted with the suffering of animals. They may respond by eating less meat, giving it up altogether, or minimizing the importance of animals. In addition, recent research by \cite{Ioa:etal:24} has shown that awareness of animal welfare issues in the dairy industry can also trigger feelings of guilt, leading to efforts to alleviate this internal conflict, similar to the responses observed with meat consumption.

In the field of computational social science, one approach to address issues such as cognitive dissonance, the behavior-intention gap, or other psycho-social phenomena that show discrepancies between private and expressed opinions (or actions) are Expressed and Private Opinions (EPOs) models \citep{Ye:etal:19,Hou:Li:Jia:21,Sha:21,Xia:Ho:Men:23,Qin:Li:Min:23,Don:etal:24}, a subset of opinion dynamics models \citep{Gra:Rus:20, Noo:20, Zha:etal:22, Li:etal:21}. These models are studied by different communities, which often use different terminology. For example, terms such as "opinions", "beliefs" and "attitudes" are often used interchangeably, despite their psychological distinctions. Similarly, in models that examine private and expressed opinions, "private" may be 
replaced by "internal" \citep{Gas:Obo:Gul:18,Ban:Olb:19} while "expressed" may be replaced by "public" \citep{Mar:08,Hua:Wen:14,Man:etal:20,Roy:Bis:21,Jac:Ban:23} or "external" \citep{Ban:Olb:19}. Such terminological variations between communities are common and not limited to opinion dynamics. However, they can pose significant challenges to interdisciplinary communication, and thus we highlight this challenge here. In this paper, we choose to use the terms "expressed opinions" and "private opinions" because they most accurately reflect the focus of our work.

EPOs models have been used to study several psycho-social phenomena, including preference falsification \citep{Leo:Ten:Mig:19}, pluralistic ignorance \citep{Cen:Wil:Mac:05, Hua:Wen:14, Ye:etal:19}, and the spiral of silence \citep{Gai:Olb:Ban:20, Cab:etal:21, Ma:Zha:21}. However, they have not been used to understand the important phenomenon of hysteresis, that occurs in many psycho-social systems, as reviewed by \cite{Szn:Jed:Kam:24}. This paper aims to fill this gap. Specifically, we ask how the mechanism of cognitive dissonance reduction affects hysteresis.

\section{Methods}
The model we consider here belongs to the broader class of Expressed-Private-Opinion (EPO) models, a general theoretical framework for studying how discrepancies between individuals' private and expressed opinions arise. Within this class of models, each agent has a private opinion, \textit{which is unknown to others but evolves under local influence from the expressed opinions of its neighbors, and an expressed opinion that varies under peer pressure to conform to the local environment} \citep{Sha:21}. 

Specifically, we introduce a modification of the model proposed by \citep{Jed:etal:18}. We study a population of $N$ agents, also referred to as voters, distributed over the vertices of a simple undirected graph of size $N$, representing a social network. If an edge connects two vertices, we call the agents at those vertices neighbors, and they can interact with each other. While different types of graphs could be used \citep{Wan:etal:20}, we focus specifically on a complete graph for this study. Despite its limitations in accurately reflecting real-world social networks, the complete graph is chosen to be consistent with the analysis of the original model, allowing for an accurate comparison by preserving the graph structure used in previous research. In addition, the complete graph is consistent with a mean-field or well-mixed population approach, which assumes homogeneity and the absence of spatial correlations among agents. This configuration implies that each pair of individuals has an equal probability of interacting, which simplifies the analytical treatment of the model. Consistent with previous studies, we will analyze the model using two methods: analytical techniques and Monte Carlo simulations.

\subsection{Model formulation}
\label{sec:model}
Agents are characterized by two dynamic binary variables: their expressed opinion, $S_k (t) = \pm 1$, and their private opinion, $\sigma_k (t) = \pm 1$, where $t$ denotes time and $k=1,\ldots,N$ indexes the agents. For brevity, we often write $S_k$ and $\sigma_k$ instead of $S_k(t)$ and $\sigma_k(t)$. We also introduce the following notation: $\uparrow$ represents an opinion of $1$, while $\downarrow$ represents $-1$. The state of an agent, defined by these two opinions, is articulated by the pair $S_k \sigma_k$. For example, $\uparrow \downarrow$ means that the voter has an expressed opinion $S_k = 1$ and a private opinion $\sigma_k = -1$. Thus, each voter can be in one of four possible states: $\uparrow \uparrow, \uparrow \downarrow, \downarrow \uparrow, \downarrow \downarrow$.

Following the original work, we use random sequential updates to simulate continuous time. In each elementary time step (update) of duration $\Delta t$, a single agent (referred to as the target) is randomly selected to reconsider its opinion. A unit of time consists of $N$ such updates, i.e. $N \Delta t = 1$. After selecting a target, we randomly select $q$ neighbors without repetitions to form a source of influence, also referred to as a $q$-panel, which will exert influence on the target. As agents cannot perceive the private opinions of others, a target is influenced solely by the expressed opinions of the selected $q$ neighbors, resulting in various possible social responses. 

\begin{figure}[h!]
    \centering
    \includegraphics[width = 0.8\textwidth]{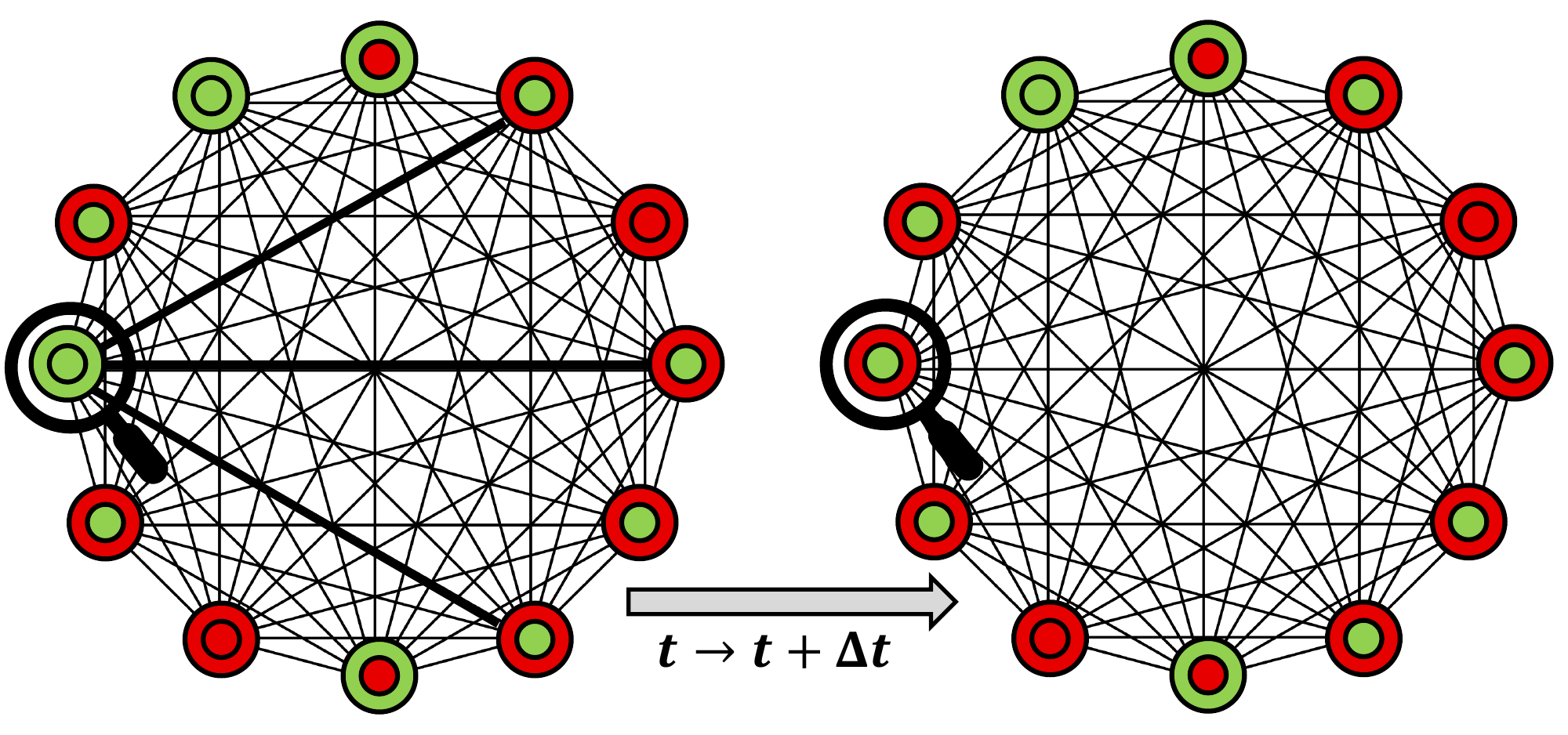}
    \caption{Example of a change in expressed opinion due to compliance with the influence group of size $q=3$, leading to dissonance. An agent under the lens is a target that updates its opinion. As a result, the number of agents in dissonance increases and the majority of expressed opinions are negative (dark red), although privately most agents are positive (light green). Implementation in NetLogo available at \href{https://barbarakaminska.github.io/q-voter-model-expressed-private/}{GitHub}. }
    \label{fig:expressed_private_ex1}
\end{figure}

\textbf{Expressed opinions may undergo changes due to:}
\begin{description}
\item[Compliance:] If a target is initially in internal harmony, i.e. $S_k = \sigma_k$, it will be less susceptible to social pressure. In this case, only a unanimous $q$-panel can influence a target, i.e., if all $q$ voters express the same opinion, a target will change its expressed opinion, even though this may lead to internal dissonance ($S_k \neq \sigma_k$), as shown in Fig. \ref{fig:expressed_private_ex1}
\item[Disinhibitory contagion:] If a target is initially in internal conflict, i.e. $S_k \neq \sigma_k$, it will be more willing to change. Therefore, it is sufficient that at least one neighbor has an opinion that is consistent with the agent's private opinion for the agent to be encouraged to express it. Thus, we replace the agent's expressed opinion with the private one if among the selected $q$ neighbors there is at least one that has expressed the same opinion as the target's private one.
\item[Independence:] If a target decides to act independently, it replaces its expressed opinion with its private one: $S_k := \sigma_k$. 
\end{description} 

\textbf{Private opinions may undergo changes due to:}
\begin{description}
\item[Conformity:] Changing the agent's private opinion also involves selecting $q$ neighbors to be the source of influence. At this point, we introduce a modification of the previous model \citep{Jed:etal:18}. Changing an agent's private opinion now requires not only the unanimous opinion of the selected $q$ neighbors, but also that the agent itself expresses the same opinion, which is shared within the $q$ panel. Consequently, the agent adjusts its private opinion to conform to the unanimous opinion expressed by the $q$ neighbors and itself. As a corollary, conformity at the private level precludes dissonance, i.e., there is no possibility of self-anticonformity or self-disagreement.
\item[Independence:] A change in private opinion as an independent decision involves rethinking a given issue, with the internal opinion changing to the opposite with a probability of 1/2. It is important to note, that according to psychological models of social response, independence implies no influence; despite social pressure, the target maintains its opinion both publicly and privately \citep{Nai:Dom:Mac:13}. Nevertheless, \cite{Jed:etal:18} assumed that even if the target is not influenced by others, it can still change its opinion through processes such as rethinking the issue, etc. To be consistent with the previous model, we will allow for such random changes here as well. 
\end{description}
 
Following \cite{Jed:etal:18}, we consider two variants of the model: (1) within AT (act then think) we update first the expressed and then the private opinion of a randomly chosen voter, while (2) within TA (think then act) we perform updates in the reverse order. The algorithm of an elementary update for the AT model is as follows:
\begin{enumerate}
    \item Choose a random voter (target), located at the site $k$. 
    \item Update expressed opinion (ACT):
    \begin{enumerate}
        \item with probability $p$ the target acts independently: $S_k := \sigma_k$   
        \item with probability $1-p$ the target is influenced by randomly chosen $q$ neighbors:
        \begin{itemize}
            \item if $S_k \neq \sigma_k$ - disinhibitory contagion: if there is at least one $S_{k_j} = \sigma_k, j=1,..,q$, then $S_k := \sigma_k$ 
            \item if $S_k = \sigma_i$ - compliance: if $S_{k_1} = ... = S_{k_q}$, then $S_k := S_{k_1}$ 
        \end{itemize}
    \end{enumerate}
    \item Update private opinion (THINK):
    \begin{enumerate}
        \item with probability $p$ the target acts independently - randomly changes opinion: $\sigma_k :=-\sigma_k$ with probability $1/2$
        \item with probability $1-p$, the target is influenced by randomly chosen $q$ neighbors, but only if they all share the same expressed opinion, and this opinion matches that of the target, i.e.: \\
        if $S_k = S_{k_1}= ... =S_{k_q}$, then $\sigma_k:= S_k$; this point distinguishes the model examined here from the model examined by \cite{Jed:etal:18}.
    \end{enumerate}
\end{enumerate}
The algorithm for the TA model differs in the order of points 2 and 3, that is, the first private opinion is~updated and after that the expressed one.
To facilitate individual exploration of the model, we have developed an implementation in NetLogo, which is available at \href{https://barbarakaminska.github.io/q-voter-model-expressed-private/}{GitHub}. This tool allows you to follow the algorithm step by step and run your own simulations. While this paper focuses primarily on the complete graph, the program also allows observation of the model on Watts-Strogatz networks.

The crucial difference between the model considered here and the model proposed by \cite{Jed:etal:18} lies in point 3(b) of the above algorithm. In the previous version of the model, it was not necessary for the public opinion of the target voter to match the public opinion of the influence group in order for the private opinion of a target to match the influence group. This means that in the previous model, the following situation was possible: an expressed and private opinion of a target voter was "YES", but the opinion of its neighbors was "NO", and as a result of social influence, the target began to privately support "NO", although in public he still claimed to support "YES". This means that social influence could lead to dissonance between private and expressed opinions. This behavior is not actually impossible and could be called strategic anticonformity. In this paper, however, we introduce a mechanism to prevent such dissonance and ask how it will affect the behavior of the model.  We do not claim that the new version of the model is more realistic than the old one, but simply ask how the mechanism that reduces cognitive dissonance at the individual level affects the behavior of the entire system. As we will see, this seemingly small change will have a significant impact on the results of the model.

\subsection{Aggregated variables}
To measure the state of the system on the macroscopic level we can use macroscopic quantities, such as the number of agents in each of four possible states $N_{\uparrow \uparrow}(t), N_{\uparrow \downarrow}(t), N_{\downarrow \uparrow}(t), N_{\downarrow \downarrow}(t)$ or corresponding intensive quantities (i.e. independent of the size of the system):
\begin{equation}
c_{\uparrow \uparrow}(t)=\frac{N_{\uparrow \uparrow}(t)}{N}, \; c_{\uparrow \downarrow}(t)=\frac{N_{\uparrow \downarrow}(t)}{N}, \; c_{\downarrow \uparrow}(t)=\frac{N_{\downarrow \uparrow}(t)}{N}, \; c_{\downarrow \downarrow}(t)=\frac{N_{\downarrow \downarrow}(t)}{N},    
\end{equation}
which we call the concentrations of a given state.
However, only 3 of them are independent, as the total number of all agents is conserved:
\begin{eqnarray}
N_{\uparrow \uparrow}(t) + N_{\uparrow \downarrow}(t) + N_{\downarrow \uparrow}(t) + N_{\downarrow \downarrow}(t) & = & N, \nonumber \\
c_{\uparrow \uparrow}(t) + c_{\uparrow \downarrow}(t) + c_{\downarrow \uparrow}(t) + c_{\downarrow \downarrow}(t) &=& 1.
\label{eq:norm}
\end{eqnarray}
We could use any $3$ of the above $4$ variables, but to get a better insight into the behavior of the system, for the presentation of the results we will use concentrations of voters with positive opinions (expressed and private) and concentrations of voters in dissonance, as done by \cite{Jed:etal:18}:\begin{eqnarray}
c_S (t) & = & c_{\uparrow \uparrow} (t) + c_{\uparrow \downarrow} (t),  \nonumber \\
c_{\sigma} (t) & = & c_{\uparrow \uparrow} (t) + c_{\downarrow \uparrow} (t),\nonumber \\
d (t) & = & c_{\uparrow \downarrow} (t) + c_{\downarrow \uparrow} (t). 
\label{eq:macro_quant}
\end{eqnarray}

We will measure the above quantities within Monte Carlo simulations, as well as calculate them analytically within the mean-field approach.

\subsection{Monte Carlo simulations}
Monte Carlo simulations are a primary tool for rigorous analysis of agent-based models. When performing Monte Carlo simulations to measure stationary values of an aggregated quantity, we prepare the initial state of the system and allow it to evolve according to predefined rules until a steady state is reached, a process often referred to as equilibration or thermalization. Once the system has reached equilibrium, we proceed to measure the quantities of interest. We typically average these measurements over time or over the ensemble of simulated states.

In the first case of time averaging, we compute the time average of the quantity of interest by summing the values of the quantity at different time steps and then dividing by the total number of measurements. In the second case, ensemble averaging, we compute the ensemble average by sampling the quantity of interest from multiple independent simulations or configurations of the system. This involves calculating the average of the quantity over all sampled states.

In simulations of this nature, time $t$ is typically measured in Monte Carlo steps or sweeps (MCS). We have defined a unit of time such that $N \Delta t = 1$ is equivalent to one MCS. Therefore, one MCS comprises $N$ elementary updates, indicating that on average, each agent undergoes one update per MCS.

In the case of a complete graph (a fixed network structure), we use time averaging. Simulations are conducted over $12 \times 10^3$ MCS, excluding the initial $10 \times 10^3$ MCS to ensure the system reaches a steady state. Analytical calculations within the mean-field approach are exact in the limit as $N \to \infty$. Therefore, to compare simulation results with analytical findings, we use sufficiently large networks. Specifically, we consider a complete graph with $N = 10^5$ nodes. As demonstrated later in this paper, these system sizes, along with the thermalization time and number of averages, yield results that closely match those obtained through analytical calculations. The errors are smaller than the size of the symbols used to represent the simulation results. 

We assume that all agents are initially in internal harmony, i.e. $S_k(0) = \sigma_k(0)$ for all $k=1,\ldots,N$. Furthermore, most of the simulations were performed for ordered initial conditions (positive consensus): $S_k(0) = \sigma_k(0) = 1$. However, we also ran a simulation with random initial conditions (positive and negative opinions were uniformly distributed among the agents), as this may affect the final state in the case where hysteresis is observed. 

\section{Calculations}
\label{sec:cal}
Since we use a random sequential update scheme, at an elementary time step $\Delta t =1/N$ the concentration $c_i$, where ${i = \uparrow\uparrow, \uparrow\downarrow, \downarrow\uparrow, \downarrow\downarrow }$ can increase or decrease by $1/N$ or remain unchanged. After \citep{Jed:etal:18} we denote the transition rates for these three events by:
\begin{eqnarray}
\gamma_i^+ & = & Prob(c_i(t + \Delta t) =  c_i (t) + 1/N), \nonumber \\
\gamma_i^- & = & Prob(c_i(t + \Delta t) =  c_i (t) - 1/N),  \nonumber \\
\gamma_i^0 & = & Prob(c_i(t + \Delta t) =  c_i (t)) = 1 - \gamma_i^+ - \gamma_i^-,
\end{eqnarray}
Actually, we do not need to define $\gamma_i^0$ because we will not use it in the calculations, but we write it here for the sake of order. The exact forms of the transition rates depend on the details of the model and are derived in subsections \ref{sec:calc_AT} and \ref{sec:calc_TA} for AT and TA models, respectively. Once we know the transition rates, we can write the evolution of the concentration of agents in a given state $i$ as follows: 
\begin{align}
    c_i(t + \Delta t) &= c_i (t) + \frac{\gamma_i^+ - \gamma_i^-}{N},
\end{align}
and in the limit $N \xrightarrow{} \infty$:
\begin{align}
    \frac{dc_i}{dt} &= \gamma_i^+ - \gamma_i^-.
\label{eq:rate_equation} 
\end{align}
We cannot solve these equations analytically, but we can solve them numerically, and we will do so to analyze the temporal behavior of the system in Sec. \ref{sec:results}.

The stationary states satisfy the condition  
\begin{align}
    \frac{dc_i}{dt} = 0, \label{eq:stationary_states}
\end{align}
which allows us to compute stationary (time-independent) values of the agent concentrations in each of the four states: $c_{\uparrow \uparrow}, c_{\uparrow \downarrow}, c_{\downarrow \uparrow}, c_{\downarrow \downarrow}$. Although the stationary concentrations of agents in each of these states are different for the AT and TA models, as will be shown in subsections \ref{sec:calc_AT} and \ref{sec:calc_TA}, it turns out that they can be reduced to the same system of two equations connecting $c_S$ and $c_\sigma$:
\begin{eqnarray}
c_{\sigma} &=& \frac{p/2+ (1-p)^2c_S^{2q}}{p+(1-p)^2\left[(1-c_s)^{2q}+c_s^{2q}\right]}, \nonumber\\
0  & = & \left[c_S-(1-p)c_S^q\right]
\left[p+(1-p)^2\left[(1-c_s)^{2q}+c_s^{2q}\right]\right] \nonumber\\
& - & \left[1-(1-p)\left[(1-c_S)^q+c_S^q\right]\right]\left[p/2+ (1-p)^2c_S^{2q}\right].  
\label{eq:final}    
\end{eqnarray}
This implies that the stationary concentrations of positive expressed and private opinions are the same for AT and TA. An analogous result, that $c_S$ and $c_{\sigma}$ are the same for AT and TA, was also obtained for the previous version of the model \citep{Jed:etal:18}. 

\subsection{Calculations for the AT model} 
\label{sec:calc_AT}
\begin{figure}[h!]
\centering
\begin{subfigure}[h]{0.73\textwidth}
\includegraphics[width=\textwidth]{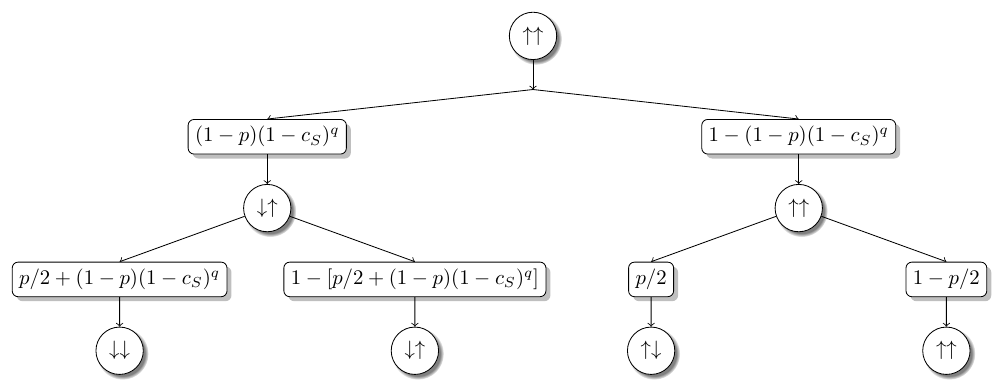}
\label{fig:AT_up_up}
\end{subfigure}
\begin{subfigure}[h]{0.73\textwidth}
\includegraphics[width=\textwidth]{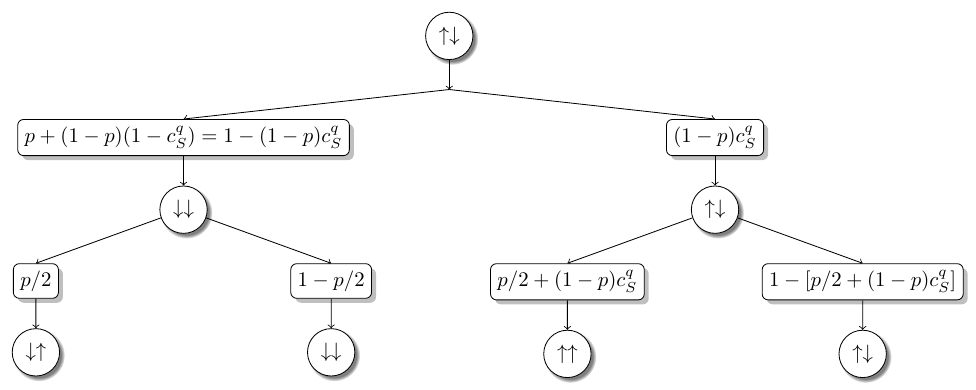}
\label{fig:AT_up_down}
\end{subfigure}
\begin{subfigure}[h]{0.73\textwidth}
\includegraphics[width=\textwidth]{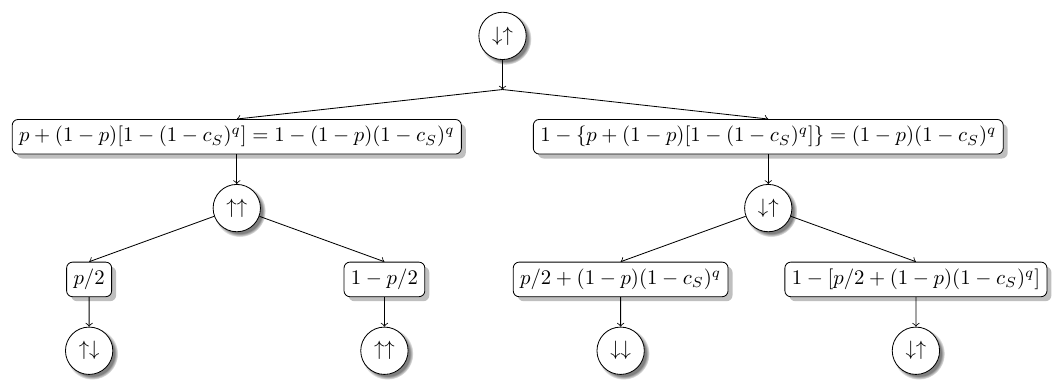}
\label{fig:AT_down_up}
\end{subfigure}
\begin{subfigure}[h]{0.73\textwidth}
\includegraphics[width=\textwidth]{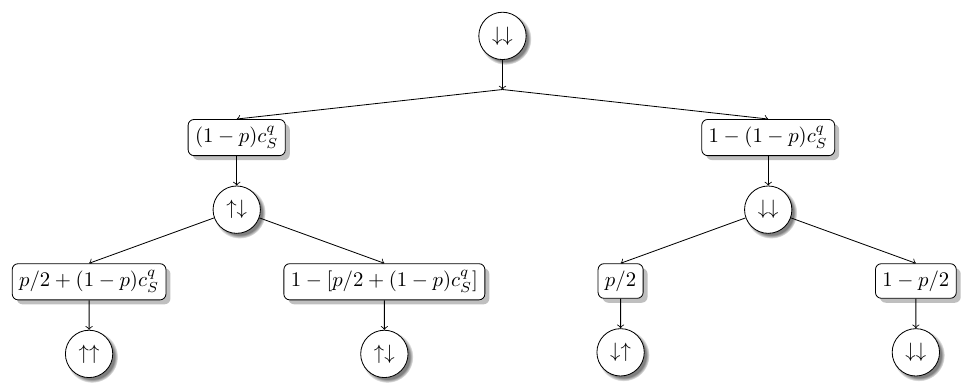}
\label{fig:AT_down_down}
\end{subfigure}
\caption{Probability tree diagrams showing the probabilities of transitions from a given state of the agent (e.g., from $\uparrow \uparrow$ in the top diagram) to all other possible states in the AT model. }
\label{fig:transitions_AT}
\end{figure}

To calculate transition rates between different states, it is convenient to draw all possible transitions in the form of trees, as shown in Fig. \ref{fig:transitions_AT}. For example, to obtain the probability of transition from $\uparrow \uparrow$ to $\downarrow \downarrow$, which corresponds to the top panel in Fig. \ref{fig:transitions_AT}, we need to multiply the product of the probabilities on the leftmost branch in the top panel of Fig. \ref{fig:transitions_AT} by $c_{\uparrow \uparrow}$, which is the probability that an agent is in the state $\uparrow \uparrow$, giving us $c_{\uparrow \uparrow}(1-p)(1-c_S)^q[p/2+(1-p)(1-c_s)^q]$. To compute $\gamma_{\uparrow \uparrow}^+$, we must compute the sum of the probabilities of transitioning to the $\uparrow \uparrow$ state from each of the other three states. Similarly, to calculate $\gamma_{\uparrow \uparrow}^-$, we need to calculate the sum of the probabilities of transitioning from state $\uparrow \uparrow$ to each of the other three states.

This gives us:
\begin{eqnarray}
\gamma_{\uparrow \uparrow}^+ &=& c_{\uparrow \downarrow}(1-p)c_S^q[p/2+(1-p)c_S^q] + c_{\downarrow \uparrow}[1-(1-p)(1-c_S)^q](1-p/2) \nonumber \\ 
& + & c_{\downarrow \downarrow}(1-p)c_S^q[p/2+(1-p)c_S^q], \nonumber \\
\gamma_{\uparrow \uparrow}^- &=& c_{\uparrow \uparrow}\{1 - [1 - (1-p)(1-c_S)^q](1 - p/2)\}.
\end{eqnarray}

From the stationary condition: 
\begin{equation}
\frac{dc_{\uparrow \uparrow}}{dt} = \gamma_{\uparrow \uparrow}^+ - \gamma_{\uparrow \uparrow}^- = 0,
\end{equation}
after simple calculations we obtain:
\begin{flalign}
    c_{\uparrow \uparrow} &=c_{\sigma} [1 - (1-p)(1-c_S)^q](1 - p/2) + (1-c_{\sigma})(1-p)c_S^q[p/2+(1-p)c_S^q]. \label{eq:c_up_up}
\end{flalign}

Using the diagrams shown in Fig. \ref{fig:transitions_AT} and repeating the analogous calculations as above, we obtain expressions for all other states: 

\begin{flalign}
    c_{\uparrow \downarrow} &= c_{\sigma}p/2[1-(1-p)(1-c_S)^q] + (1-c_{\sigma})(1-p)c_S^q[1-p/2-(1-p)c_S^q], \label{eq:c_up_down} \\
    \nonumber \\    
    c_{\downarrow \uparrow} &= c_{\sigma}(1-p)(1-c_S)^q[1-p/2-(1-p)(1-c_s)^q] + (1-c_{\sigma})[1-(1-p)c_S^q]p/2,  \label{eq:c_down_up}\\ 
    \nonumber \\
    c_{\downarrow \downarrow} &= c_{\sigma}(1-p)(1-c_S)^q[p/2+(1-p)(1-c_s)^q] + (1-c_{\sigma})[1-(1-p)c_S^q](1-p/2).  \label{eq:c_down_down}
\end{flalign}

Next, by adding \eqref{eq:c_up_up} and \eqref{eq:c_up_down} we obtain an equation for $c_S$ that contains only the terms: $c_S, c_{\sigma}$:
\begin{flalign}
c_S &= c_{\uparrow \uparrow} + c_{\uparrow \downarrow} = c_{\sigma}[1-(1-p)[(1-c_S)^q+c_S^q]]+(1-p)c_S^q.  \label{eq:AT_c_S}
\end{flalign}
Similarly to obtain an equation for $c_{\sigma}$ we add \eqref{eq:c_up_up} and \eqref{eq:c_down_up}:
\begin{flalign}
c_{\sigma} &= c_{\uparrow \uparrow} + c_{\downarrow \uparrow} = c_{\sigma} =c_{\sigma}(1-p/2)  +(1-c_{\sigma})(1-p)^2c_S^{2q}-c_{\sigma}(1-p)^2(1-c_s)^{2q} + (1-c_{\sigma})p/2. 
\end{flalign}

This allows us to get $c_{\sigma}$ as a function of $c_S$
\begin{flalign}
     c_{\sigma} &= \frac{p/2+ (1-p)^2c_S^{2q}}{p+(1-p)^2[(1-c_s)^{2q}+c_s^{2q}]}.
\label{eq:AT_c_sigma}
\end{flalign}

In order to obtain an equation with only one unknown  $c_S$ we can insert \eqref{eq:AT_c_sigma} into \eqref{eq:AT_c_S}: 
\begin{flalign}
[c_S-(1-p)c_S^q][p+(1-p)^2[(1-c_s)^{2q}+c_s^{2q}]] 
=  [1-(1-p)[(1-c_S)^q+c_S^q]][p/2+ (1-p)^2c_S^{2q}].  \label{eq:final_AT}
\end{flalign}
Eqs. (\ref{eq:AT_c_S})--(\ref{eq:final_AT}) give the set of Eqs. (\ref{eq:final}), which is used to obtain results, that are presented in Sec. \ref{sec:results}.
To calculate the dissonance we add  Eq. \eqref{eq:c_up_down} and Eq. \eqref{eq:c_down_up}:\begin{flalign}
 d &= c_{\uparrow \downarrow} + c_{\downarrow \uparrow} = p/2 + c_{\sigma}(1-p)^2(1-c_S)^q[1-(1-c_S)^q] + (1-c_{\sigma})(1-p)^2c_S^q[1-c_S^q]. \label{eq:AT_dissonance}
\end{flalign}

\subsection{Calculations for the TA model} 
\label{sec:calc_TA}
\begin{figure}[h!]
\centering
\begin{subfigure}[h]{0.73\textwidth}
\includegraphics[width=\textwidth]{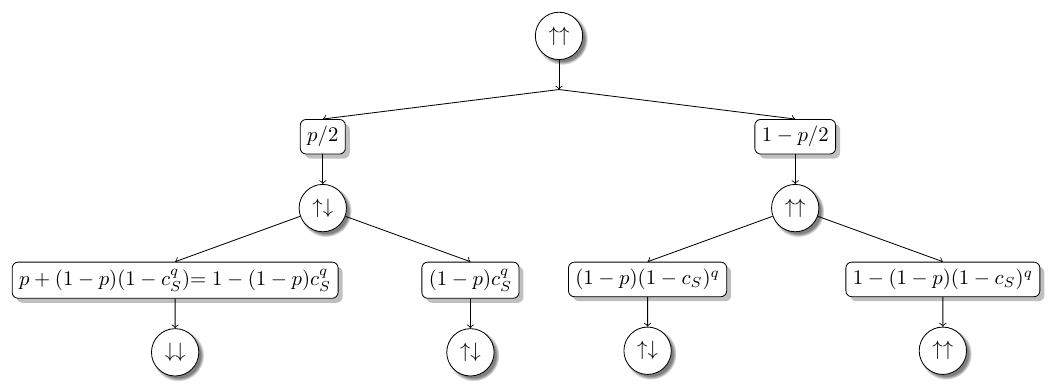}
\label{fig:TA_up_up}
\end{subfigure}
\begin{subfigure}[h]{0.73\textwidth}
\includegraphics[width=\textwidth]{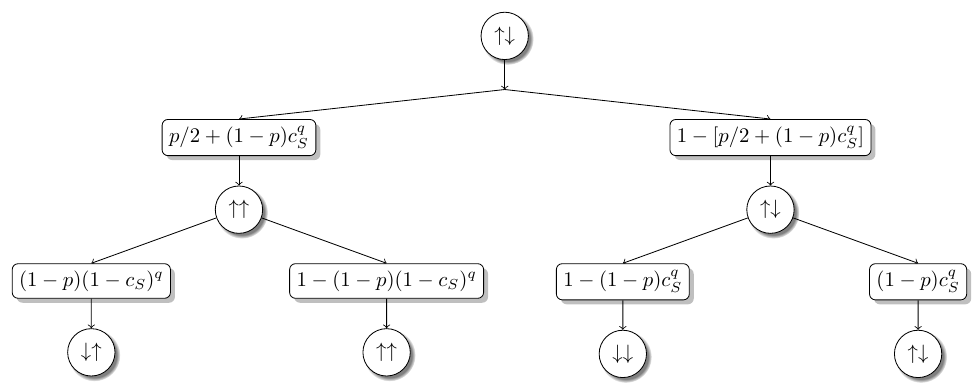}
\label{fig:TA_up_down}
\end{subfigure}
\begin{subfigure}[h]{0.73\textwidth}
\includegraphics[width=\textwidth]{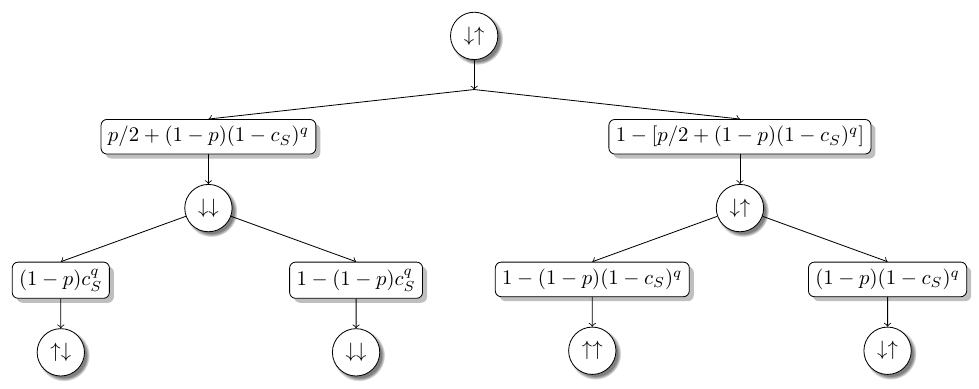}
\label{fig:TA_down_up}
\end{subfigure}
\begin{subfigure}[h]{0.73\textwidth}
\includegraphics[width=\textwidth]{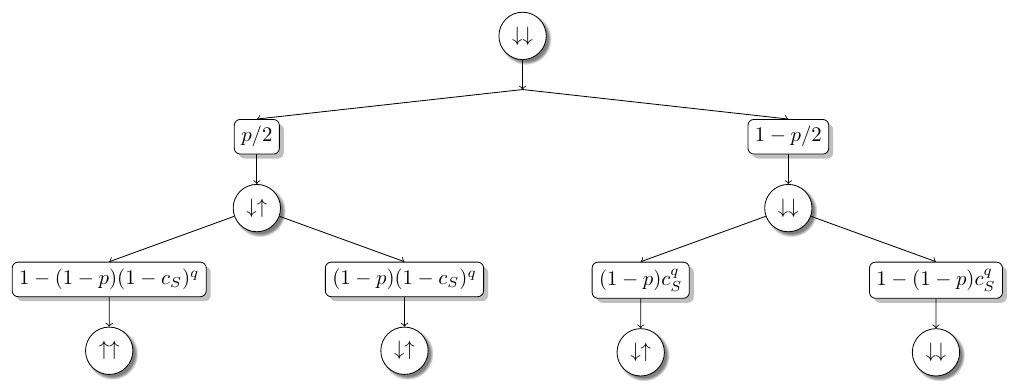}
\label{fig:TA_down_down}
\end{subfigure}
\caption{Probability tree diagrams showing the probabilities of transitions from a given state of the agent (e.g., from $\uparrow \uparrow$ in the top diagram) to all other possible states in the TA model. }
\label{fig:transitions_TA}
\end{figure}

As in the case of the AT model, we can use the diagrams shown in Figs. \ref{fig:transitions_TA} to write down all transition rates, for example:
\begin{flalign}
\gamma_{\uparrow \uparrow}^+ =& c_{\uparrow \downarrow}[p/2+(1-p)c_S^q][1 - (1-p)(1-c_S)^q]  \nonumber \\&+ c_{\downarrow \uparrow}[1-p/2-(1-p)(1-c_S)^q][1 - (1-p)(1-c_S)^q] \nonumber \\&+ c_{\downarrow \downarrow}p/2[1 - (1-p)(1-c_S)^q], \\
\gamma_{\uparrow \uparrow}^- =& c_{\uparrow \uparrow}[1-(1 - p/2)[1 - (1-p)(1-c_S)^q] ],
\end{flalign}
which gives us:
\begin{flalign}
    c_{\uparrow \uparrow} &= c_{\uparrow \uparrow} (1 - p/2)[1 - (1-p)(1-c_S)^q] + c_{\uparrow \downarrow}[p/2+(1-p)c_S^q][1 - (1-p)(1-c_S)^q] \nonumber \\ &+ c_{\downarrow \uparrow}[1-p/2-(1-p)(1-c_S)^q][1 - (1-p)(1-c_S)^q] + c_{\downarrow \downarrow}p/2[1 - (1-p)(1-c_S)^q]. \label{eq:TA_c_up_up}
\end{flalign}

We proceed analogously with the remaining states and obtain the following equations:
\begin{flalign}
    c_{\uparrow \downarrow} &= c_{\uparrow \uparrow}p/2(1-p)c_S^q + c_{\uparrow \downarrow}[1-p/2-(1-p)c_S^q](1-p)c_S^q \nonumber  \\ &+ c_{\downarrow \uparrow}[p/2+(1-p)(1-c_S)^q](1-p)c_S^q + c_{\downarrow \downarrow}(1-p/2)(1-p)c_S^q, \label{eq:TA_c_up_down} \\ 
    \nonumber  \\
    c_{\downarrow \uparrow} &= c_{\uparrow \uparrow}(1-p/2)(1-p)(1-c_S)^q + c_{\uparrow \downarrow}[p/2+(1-p)c_S^q](1-p)(1-c_S)^q \nonumber  \\&+ c_{\downarrow \uparrow}[1-p/2-(1-p)(1-c_S)^q](1-p)(1-c_S)^q + c_{\downarrow \downarrow}p/2(1-p)(1-c_S)^q,  \label{eq:TA_c_down_up} \\ 
    \nonumber \\
    c_{\downarrow \downarrow} &= c_{\uparrow \uparrow}p/2[1-(1-p)c_S^q] + c_{\uparrow \downarrow}[1-p/2-(1-p)c_S^q][1-(1-p)c_S^q] \nonumber \\& + c_{\downarrow \uparrow}[p/2+(1-p)(1-c_S)^q][1-(1-p)c_S^q] + c_{\downarrow \downarrow}(1-p/2)[1-(1-p)c_S^q]. \label{eq:TA_c_down_down}
\end{flalign}

Unlike in the AT model, after some calculations we can obtained simplified forms of Eqs. \eqref{eq:TA_c_up_up}-\eqref{eq:TA_c_down_down}: 
\begin{flalign}
c_{\uparrow \uparrow} &= c_{\sigma}[1 - (1-p)(1-c_S)^q], \\
c_{\uparrow \downarrow } &= (1-c_{\sigma})(1-p)c_S^q,  \\
c_{\downarrow \uparrow} &= c_{\sigma}(1-p)(1-c_S)^q, \\
c_{\downarrow \downarrow} &= (1-c_{\sigma})[1-(1-p)c_S^q].
\end{flalign}
Although these formulas are much simpler than in the AT model, after performing appropriate summations we obtain the same system of equations for $c_S$ and $c_\sigma$, given by Eq. (\ref{eq:final}), as in the previous version of the model.
Only the formula for the dissonance is indeed much simpler and different than for the AT
\begin{flalign}
d  &= (1-p)[(1-c_{\sigma})c_S^q + c_{\sigma}(1-c_S)^q]. \label{eq:TA_dissonance} 
\end{flalign}
The explanation for this difference in dissonance between the two versions of the model, which was strongly emphasized by \cite{Jed:etal:18}, will be presented in  Section \ref{sec:diss}.

\section{Results}
\label{sec:results}
Let us start with an example illustrating the time evolution of the concentration of positive opinions (expressed and private) and dissonance, defined by Eqs. \eqref{eq:macro_quant}. The temporal behavior of the system can be derived by numerically solving the system of Eqs. \eqref{eq:rate_equation} and by Monte Carlo simulations. The results of these two approaches, shown in Fig. \ref{fig:trajectory_q_3}, provide several pieces of information:

\begin{figure}[h!]
    \centering
    \includegraphics[width=\textwidth]{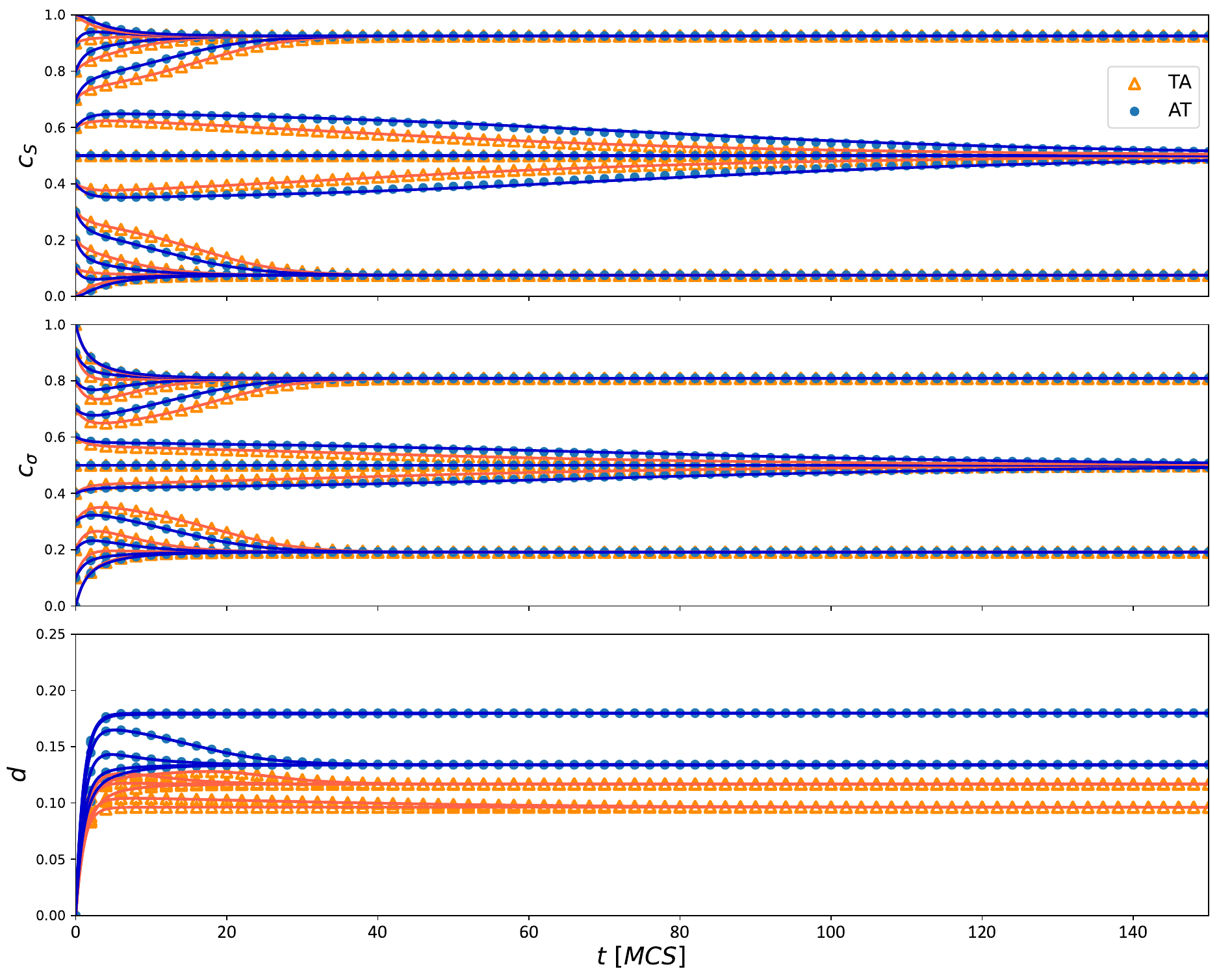}
    \caption{\textbf{An example of the time evolution of the concentrations of positive expressed (top panel) and private (middle panel) opinions, as well as dissonance (bottom panel), obtained for $q = 3$ and $p = 0.23$. } Solid lines represent numerical solutions of the equations \eqref{eq:rate_equation}, and symbols represent trajectories obtained from Monte Carlo simulations performed on a system of size $N = 10^5$ nodes and averaged over 100 samples. }
    \label{fig:trajectory_q_3}
\end{figure}

\begin{itemize}
    \item They confirm that for a complete graph of sufficient size $N$ (here $N=10^5$) the Monte Carlo simulation results agree with the analytical ones.
    \item Concentrations of positive opinions $c_S$ and $c_{\sigma}$ differ for AT and TA models only in the early stages of evolution, eventually converging to the same stationary states, consistent with conclusions from analytical calculations.
    \item The dissonance $d$ is different for AT and TA, which is also consistent with the analytical results presented in sections \ref{sec:calc_AT} and \ref{sec:calc_TA}. We will comment on this result in more detail later in section \ref{sec:diss}, as it will turn out that there is, in a sense, an apparent difference between the AT and TA models that is only due to the timing of the measurement.
    \item At both levels (expressed and private), one of two types of stationary states can be reached: agreement or disagreement. We define agreement as a state in which one opinion dominates the other. On the expressed level it corresponds to $c_S \ne 1/2$, and on the private level it corresponds to $c_{\sigma} \ne 1/2$. Conversely, disagreement means $c_S = 1/2$ on the expressed level and $c_{\sigma} = 1/2$ on the private level. The steady state depends on the initial state: for the same model parameters $p$ and $q$, disagreement or agreement can be reached. This result, called hysteresis, will be discussed later.
\end{itemize}

Let us now focus on the stationary states, which are shown in Fig. \ref{fig:stationary_states_proper}. The dependence between the stationary values of the macroscopic quantities and the probability of independence $p$ reveals the existence of the regime shift (phase transition) between agreement and disagreement. There is a critical value $p=p^*_{low}$, depending on the size of the influence group $q$, below which agreement is reached, independent of the initial state of the system. On the other hand, there is a critical value $p=p^*_{up}$, again depending on the size of the influence group $q$, above which disagreement is always reached. Finally, there is a range $p \in (p^*_{low},p^*_{up})$ in which hysteresis occurs, i.e., depending on the initial state, agreement or disagreement is reached. For small sizes of the influence group $q=1$ and $q=2$ we see that $p=p^*_{low}=p^*_{up}=p^*$, so there is no hysteresis. However, it does appear for $q \ge 3$. For all values of $q$, the critical values $p=p^*_{low},p^*_{up}$ are the same for expressed and private opinions. However, the overall dependence on the value of the parameter $p$ is different for private and expressed opinions when only the size of the influence group is $q>1$. Specifically, as can be seen in Fig. \ref{fig:stationary_states_proper}:

\begin{figure}[h]
    \centering
    \includegraphics[width=\textwidth]{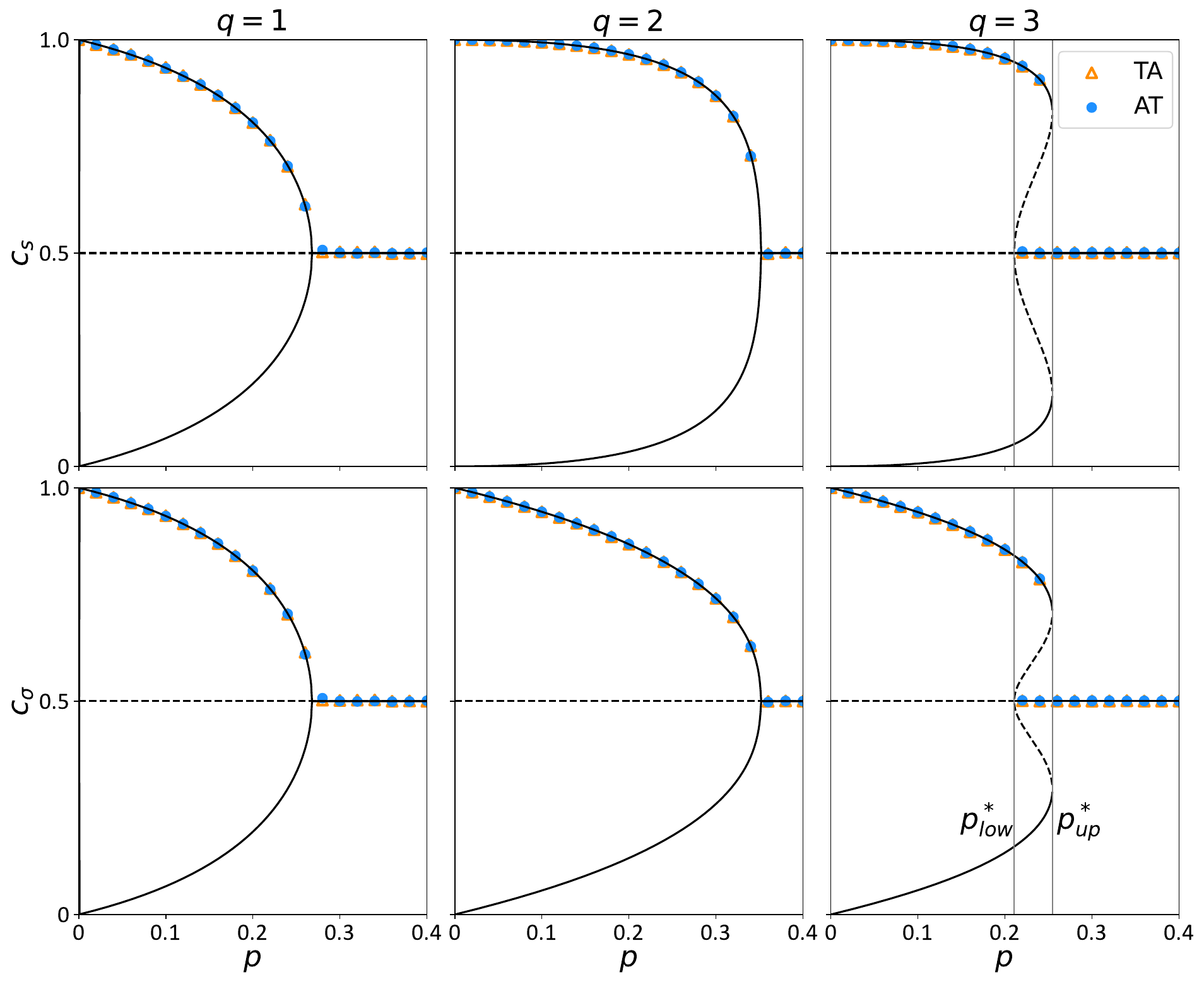}
    \caption{\textbf{Stationary concentrations of positive opinions at public (upper panel) and private level (bottom panel) as a function of the probability of independence $p$.} Columns correspond to different sizes of influence group $q=1$ (left), $q= 2$ (middle) and $q= 3$ (right). Symbols represent results of Monte Carlo simulations (triangles for TA model and circles for AT model). Lines represent analytical results. }
    \label{fig:stationary_states_proper}
\end{figure}

\begin{itemize}
    \item For $q=1$, which corresponds to the so-called linear voter model (often just called a voter model), the dependence between the concentration of positive opinions and the parameter $p$ is identical for expressed and private opinions, and there is no hysteresis.
    \item For $q=2$ there is still no hysteresis, i.e. $p=p^*_{low}=p^*_{up}=p^*$, but the dependence between the concentration of positive opinions and the parameter $p$ is different for expressed and private opinions. For $p<p^*$ the agreement is stronger, i.e. there is a greater dominance of one opinion over the other, at the expressed level than at the private level. This result is reminiscent of the phenomenon of the spiral of silence.
    \item For $q=3$ hysteresis occurs, which means that for $p \in (p^*_{low},p^*_{up})$ the final state depends on the initial state. Similarly, as for $q=2$, in case of agreement, there is a greater dominance of one opinion over the other, at the expressed level than at the private level. 
\end{itemize}

\begin{figure}[h]
    \centering
    \includegraphics[width=\textwidth]{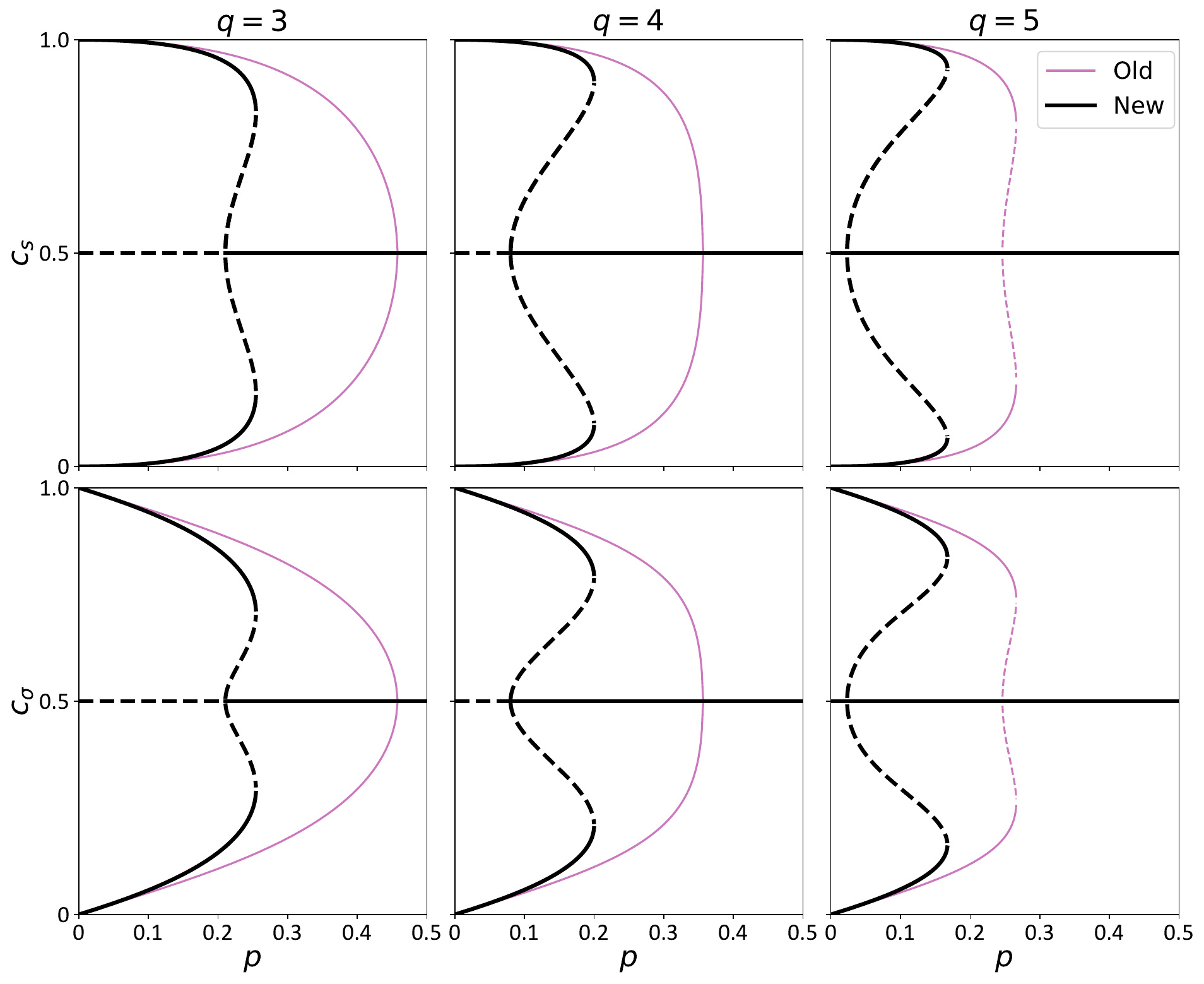}
    \caption{\textbf{Comparison between the new model presented here and the previous (old) model by \cite{Jed:etal:18}: Stationary concentrations of positive opinions at the public (top panel) and private (bottom panel) levels as a function of the probability of independence $p$}. The columns correspond to different sizes of the influence group $q=3$ (left), $q=4$ (middle), and $q=5$ (right). Black bold lines represent the new model proposed in this paper, while purple lines represent results from the old model proposed by \cite{Jed:etal:18}. }
    \label{fig:comparison}
\end{figure}

It is worth noting that in the old version of the model proposed by \cite{Jed:etal:18} there was no agreement at all for $q=1$, i.e. for any value of $p>0$ the steady state was disagreement. Furthermore, hysteresis occurred only for $q \le 5$. To clearly see the difference between the model analyzed here and the previous one, we have prepared Fig. \ref{fig:comparison}. It can be seen that not only the new version of the model supports hysteresis (i.e. it appears already for $q \le 3$), but in addition the transition between agreement and disagreement occurs for smaller values of $p$, i.e. for a wider range of $p$ there is disagreement. 

\subsection{Dissonance}
\label{sec:diss}
In Fig. \ref{fig:trajectory_q_3} it can be seen that the stationary concentrations of agents in dissonance differentiate the two variants of the model -- they are larger for the AT (act then think) model. We could draw the nice conclusion that it is better to think first to avoid dissonance. Indeed, such an observation has been made by
\cite{Jed:etal:18}, where the original model was introduced. However, no explanation was given for this result, and it was somewhat puzzling why the order of updating (AT vs. TA) did not affect opinions but dissonance. Here, we tried to better understand the effect of the updating scheme on dissonance, and this led us to the solution of this puzzle. It turns out that only the moment of measurement matters, as shown in Fig. \ref{fig:dissonance_measurment}. More details are given below.

Until now, all averaged quantities were measured after the entire elementary update, which consists of updating both the expressed opinion (ACT) and the private opinion (THINK). That is, depending on the version of the model, the measurement was made after THINK (for the AT model) or after ACT (for the TA model). To understand the source of the differences between the models, we decide to change the way we measure the averaged quantities. Namely, for both versions of the model, we make two measurements for each update - one after ACT and the other after THINK. Then, as always we average the results. As can be seen in Fig. \ref{fig:dissonance_measurment}, the level of dissonance depends on the last step performed (whether it was an update of public or private opinion) and not on the version of the model.

 \begin{figure}[h]
    \centering
    \includegraphics[width=0.65\textwidth]{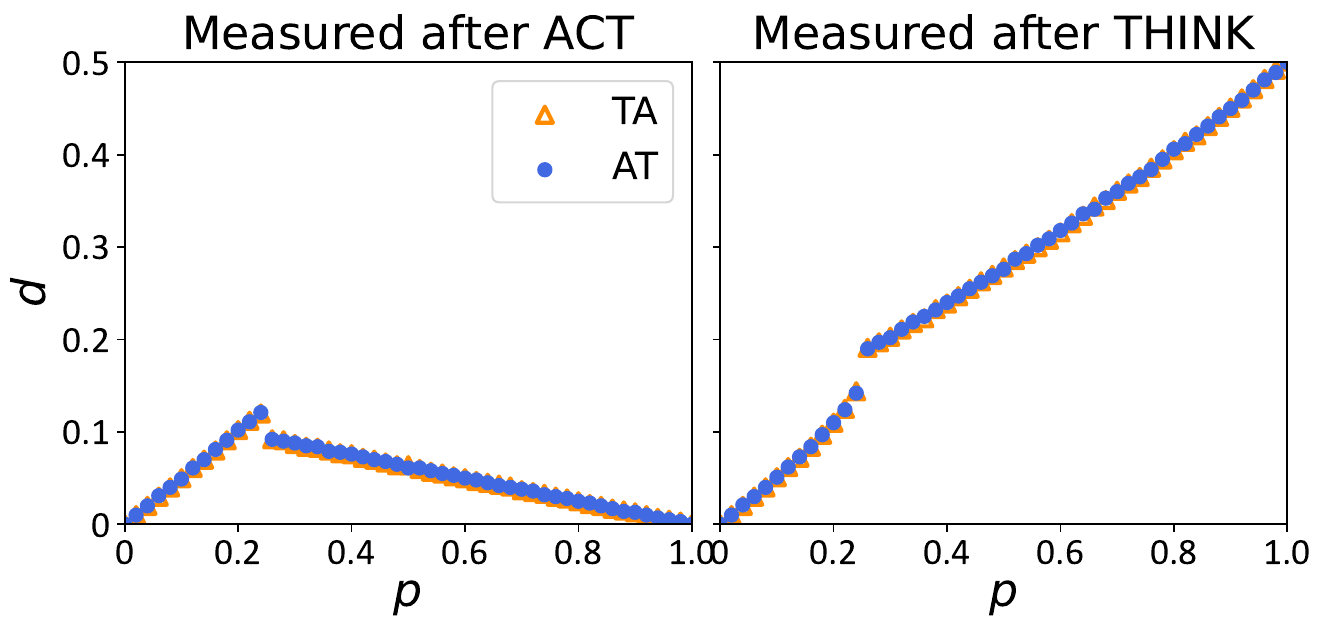}
    \caption{Stationary concentration of agents in dissonance measured after update of public (ACT) or private (THINK) opinion in AT and TA variants of the model. }
    \label{fig:dissonance_measurment}
\end{figure}

\section{Conclusions}
In recent years, there has been increased interest in the study of collective adaptation, especially in the context of complex and rapidly changing social, environmental, and technological landscapes \citep{Gal:etal:23}. A major challenge to collective adaptation is social hysteresis - a phenomenon in which a system's response to external changes is delayed and influenced by its past state (emergent collective memory); for a recent review see \citep{Szn:Jed:Kam:24}. Understanding the factors that contribute to hysteresis is therefore crucial.

To our knowledge, the question of how cognitive dissonance affects social hysteresis has not been addressed. Therefore, our work aimed to fill this gap by investigating the influence of cognitive dissonance on hysteresis using an agent-based model of Expressed and Private Opinions (EPOs) originally introduced by \cite{Jed:etal:18}. In our study, we modified this model by incorporating a behavioral rule: private opinions cannot change due to conformity to the influence group if doing so would create internal dissonance, i.e., a discrepancy between an individual's expressed and private opinions. This modification introduces an additional mechanism for reducing cognitive dissonance.

This seemingly small change at the individual level led to the emergence of several collective phenomena. In particular, hysteresis was observed for influence group sizes $q \ge 3$, whereas in the absence of cognitive dissonance reduction, hysteresis was evident only for larger influence group sizes ($q \ge 5$), as shown in Fig. \ref{fig:comparison}. In addition, the regime shift between agreement (or adoption of a behavior) and disagreement occurred within a narrow parameter range $p \in (0,p^*)$, whereas without the dissonance reduction mechanism the range of $p$ allowing agreement was much wider due to a significantly higher value of $p^*$.

It is important to note that cognitive dissonance reduction was already present in the previous version of the model through disinhibitory contagion, where dissonance was reduced by externalizing private opinions through verbalization or action. In our study, we retain this mechanism while adding a new one that prevents individuals from acting differently than they think. Our findings suggest that reducing cognitive dissonance may hinder communication and, furthermore, also promote social hysteresis.

The behavior we have prohibited in the current version of the model is known as strategic self-anticonformity, also called reverse psychology \citep{Mac:Nai:Har:11,Nai:Dom:Mac:13}. This behavior involves expressing opinions or behaving differently than others, even when our personal opinions coincide with theirs. Reverse psychology has been empirically demonstrated in social experiments and has been used by marketers for decades to influence consumer behavior. However, it has not been linked to consensus or social hysteresis.

Based on our research, we argue that strategic anticonformity can be beneficial in achieving consensus and preventing social hysteresis. A similar conclusion regarding strategic anticonformity was recently reached in another agent-based model by \cite{Lip:Szn:22}. It is important to note that \cite{Lip:Szn:22} considered only expressed opinions in a three-state model with additional bounded confidence, while our model considers both private and expressed opinions in a two-state format. Despite these differences, both studies show that strategic anticonformity can facilitate consensus building and mitigate social polarization.

We began our work by noting that in reality, as demonstrated by a number of social experiments, there is a natural tendency to reduce cognitive dissonance. We therefore introduced such a mechanism into our model. However, our research has shown that such a mechanism is not beneficial for the entire social group. Therefore, strategic anticonformity (getting rid of the new mechanism introduced here) may be a good policy that not only facilitates reaching consensus, but also counteracts hysteresis, which delays the system's response to external changes.

\section{Acknowledgment}
This research was funded by National Science Centre
Poland Grant 2019/35/B/HS6/02530. The authors thank Arkadiusz J\c{e}drzejewski and Arkadiusz Lipiecki for helpful discussions.

\printcredits

\bibliographystyle{cas-model2-names}

\bibliography{EPO}

\begin{thebibliography}{32}
\expandafter\ifx\csname natexlab\endcsname\relax\def\natexlab#1{#1}\fi
\providecommand{\url}[1]{\texttt{#1}}
\providecommand{\href}[2]{#2}
\providecommand{\path}[1]{#1}
\providecommand{\DOIprefix}{doi:}
\providecommand{\ArXivprefix}{arXiv:}
\providecommand{\URLprefix}{URL: }
\providecommand{\Pubmedprefix}{pmid:}
\providecommand{\doi}[1]{\href{http://dx.doi.org/#1}{\path{#1}}}
\providecommand{\Pubmed}[1]{\href{pmid:#1}{\path{#1}}}
\providecommand{\bibinfo}[2]{#2}
\ifx\xfnm\relax \def\xfnm[#1]{\unskip,\space#1}\fi
\bibitem[{Banisch and Olbrich(2018)}]{Ban:Olb:19}
\bibinfo{author}{Banisch, S.}, \bibinfo{author}{Olbrich, E.}, \bibinfo{year}{2018}.
\newblock \bibinfo{title}{Opinion polarization by learning from social feedback}.
\newblock \bibinfo{journal}{The Journal of Mathematical Sociology} \bibinfo{volume}{43}, \bibinfo{pages}{76–103}.
\newblock \DOIprefix\doi{10.1080/0022250x.2018.1517761}.
\bibitem[{Cabrera et~al.(2021)Cabrera, Ross, R\"{o}chert, Br\"{u}nker and Stieglitz}]{Cab:etal:21}
\bibinfo{author}{Cabrera, B.}, \bibinfo{author}{Ross, B.}, \bibinfo{author}{R\"{o}chert, D.}, \bibinfo{author}{Br\"{u}nker, F.}, \bibinfo{author}{Stieglitz, S.}, \bibinfo{year}{2021}.
\newblock \bibinfo{title}{The influence of community structure on opinion expression: an agent-based model}.
\newblock \bibinfo{journal}{Journal of Business Economics} \bibinfo{volume}{91}, \bibinfo{pages}{1331--1355}.
\newblock \DOIprefix\doi{10.1007/s11573-021-01064-7}.
\bibitem[{Centola et~al.(2005)Centola, Willer and Macy}]{Cen:Wil:Mac:05}
\bibinfo{author}{Centola, D.}, \bibinfo{author}{Willer, R.}, \bibinfo{author}{Macy, M.}, \bibinfo{year}{2005}.
\newblock \bibinfo{title}{The emperor's dilemma: A computational model of self-enforcing norms}.
\newblock \bibinfo{journal}{American Journal of Sociology} \bibinfo{volume}{110}, \bibinfo{pages}{1009--1040}.
\newblock \DOIprefix\doi{10.1086/427321}.
\bibitem[{Dong et~al.(2024)Dong, Hu, Zhao and Peng}]{Don:etal:24}
\bibinfo{author}{Dong, J.}, \bibinfo{author}{Hu, J.}, \bibinfo{author}{Zhao, Y.}, \bibinfo{author}{Peng, Y.}, \bibinfo{year}{2024}.
\newblock \bibinfo{title}{Opinion formation analysis for expressed and private opinions (epos) models: Reasoning private opinions from behaviors in group decision-making systems}.
\newblock \bibinfo{journal}{Expert Systems with Applications} \bibinfo{volume}{236}, \bibinfo{pages}{121292}.
\newblock \DOIprefix\doi{10.1016/j.eswa.2023.121292}.
\bibitem[{Festinger(1957)}]{fes:57}
\bibinfo{author}{Festinger, L.}, \bibinfo{year}{1957}.
\newblock \bibinfo{title}{A theory of cognitive dissonance}.
\newblock \bibinfo{publisher}{Stanford University Press}.
\bibitem[{Gaisbauer et~al.(2020)Gaisbauer, Olbrich and Banisch}]{Gai:Olb:Ban:20}
\bibinfo{author}{Gaisbauer, F.}, \bibinfo{author}{Olbrich, E.}, \bibinfo{author}{Banisch, S.}, \bibinfo{year}{2020}.
\newblock \bibinfo{title}{Dynamics of opinion expression}.
\newblock \bibinfo{journal}{Physical Review E} \bibinfo{volume}{102}.
\newblock \DOIprefix\doi{10.1103/physreve.102.042303}.
\bibitem[{Galesic et~al.(2023)Galesic, Barkoczi, Berdahl, Biro, Carbone et~al.}]{Gal:etal:23}
\bibinfo{author}{Galesic, M.}, \bibinfo{author}{Barkoczi, D.}, \bibinfo{author}{Berdahl, A.M.}, \bibinfo{author}{Biro, D.}, \bibinfo{author}{Carbone, G.}, et~al., \bibinfo{year}{2023}.
\newblock \bibinfo{title}{Beyond collective intelligence: Collective adaptation}.
\newblock \bibinfo{journal}{Journal of the Royal Society Interface} \bibinfo{volume}{20}, \bibinfo{pages}{20220736}.
\newblock \DOIprefix\doi{10.1098/RSIF.2022.0736}.
\bibitem[{Gastner et~al.(2018)Gastner, Oborny and Guly{\'{a}}s}]{Gas:Obo:Gul:18}
\bibinfo{author}{Gastner, M.T.}, \bibinfo{author}{Oborny, B.}, \bibinfo{author}{Guly{\'{a}}s, M.}, \bibinfo{year}{2018}.
\newblock \bibinfo{title}{Consensus time in a voter model with concealed and publicly expressed opinions}.
\newblock \bibinfo{journal}{Journal of Statistical Mechanics: Theory and Experiment} \bibinfo{volume}{2018}, \bibinfo{pages}{063401}.
\newblock \DOIprefix\doi{10.1088/1742-5468/aac14a}.
\bibitem[{Grabisch and Rusinowska(2020)}]{Gra:Rus:20}
\bibinfo{author}{Grabisch, M.}, \bibinfo{author}{Rusinowska, A.}, \bibinfo{year}{2020}.
\newblock \bibinfo{title}{A survey on nonstrategic models of opinion dynamics}.
\newblock \bibinfo{journal}{Games} \bibinfo{volume}{11}, \bibinfo{pages}{65}.
\newblock \DOIprefix\doi{10.3390/g11040065}.
\bibitem[{Hou et~al.(2021)Hou, Li and Jiang}]{Hou:Li:Jia:21}
\bibinfo{author}{Hou, J.}, \bibinfo{author}{Li, W.}, \bibinfo{author}{Jiang, M.}, \bibinfo{year}{2021}.
\newblock \bibinfo{title}{Opinion dynamics in modified expressed and private model with bounded confidence}.
\newblock \bibinfo{journal}{Physica A: Statistical Mechanics and its Applications} \bibinfo{volume}{574}, \bibinfo{pages}{125968}.
\newblock \DOIprefix\doi{10.1016/j.physa.2021.125968}.
\bibitem[{Huang and Wen(2014)}]{Hua:Wen:14}
\bibinfo{author}{Huang, C.Y.}, \bibinfo{author}{Wen, T.H.}, \bibinfo{year}{2014}.
\newblock \bibinfo{title}{A novel private attitude and public opinion dynamics model for simulating pluralistic ignorance and minority influence}.
\newblock \bibinfo{journal}{Journal of Artificial Societies and Social Simulation} \bibinfo{volume}{17}.
\newblock \DOIprefix\doi{10.18564/jasss.2517}.
\bibitem[{Ioannidou et~al.(2024)Ioannidou, Lesk, Stewart-Knox and Francis}]{Ioa:etal:24}
\bibinfo{author}{Ioannidou, M.}, \bibinfo{author}{Lesk, V.}, \bibinfo{author}{Stewart-Knox, B.}, \bibinfo{author}{Francis, K.B.}, \bibinfo{year}{2024}.
\newblock \bibinfo{title}{Don't mind milk? the role of animal suffering, speciesism, and guilt in the denial of mind and moral status of dairy cows}.
\newblock \bibinfo{journal}{Food Quality and Preference} \bibinfo{volume}{114}.
\newblock \DOIprefix\doi{10.1016/j.foodqual.2023.105082}.
\bibitem[{Jacob and Banisch(2023)}]{Jac:Ban:23}
\bibinfo{author}{Jacob, D.}, \bibinfo{author}{Banisch, S.}, \bibinfo{year}{2023}.
\newblock \bibinfo{title}{Polarization in social media: A virtual worlds-based approach}.
\newblock \bibinfo{journal}{Journal of Artificial Societies and Social Simulation} \bibinfo{volume}{26}, \bibinfo{pages}{11}.
\newblock \DOIprefix\doi{10.18564/jasss.5170}.
\bibitem[{Juvan and Dolnicar(2014)}]{Juv:Dol:14}
\bibinfo{author}{Juvan, E.}, \bibinfo{author}{Dolnicar, S.}, \bibinfo{year}{2014}.
\newblock \bibinfo{title}{The attitude-behaviour gap in sustainable tourism}.
\newblock \bibinfo{journal}{Annals of Tourism Research} \bibinfo{volume}{48}, \bibinfo{pages}{76 – 95}.
\newblock \DOIprefix\doi{10.1016/j.annals.2014.05.012}.
\bibitem[{Jędrzejewski et~al.(2018)Jędrzejewski, Marcjasz, Nail and Sznajd-Weron}]{Jed:etal:18}
\bibinfo{author}{Jędrzejewski, A.}, \bibinfo{author}{Marcjasz, G.}, \bibinfo{author}{Nail, P.R.}, \bibinfo{author}{Sznajd-Weron, K.}, \bibinfo{year}{2018}.
\newblock \bibinfo{title}{Think then act or act then think?}
\newblock \bibinfo{journal}{PLOS ONE} \bibinfo{volume}{13}, \bibinfo{pages}{1--19}.
\newblock \DOIprefix\doi{10.1371/journal.pone.0206166}.
\bibitem[{Le{\'{o}}n-Medina et~al.(2019)Le{\'{o}}n-Medina, Tena-S{\'{a}}nchez and Miguel}]{Leo:Ten:Mig:19}
\bibinfo{author}{Le{\'{o}}n-Medina, F.J.}, \bibinfo{author}{Tena-S{\'{a}}nchez, J.}, \bibinfo{author}{Miguel, F.J.}, \bibinfo{year}{2019}.
\newblock \bibinfo{title}{Fakers becoming believers: how opinion dynamics are shaped by preference falsification, impression management and coherence heuristics}.
\newblock \bibinfo{journal}{Quality \& Quantity} \bibinfo{volume}{54}, \bibinfo{pages}{385--412}.
\newblock \DOIprefix\doi{10.1007/s11135-019-00909-2}.
\bibitem[{Li et~al.(2021)Li, Liu, Cao, Wang and Zhang}]{Li:etal:21}
\bibinfo{author}{Li, Y.}, \bibinfo{author}{Liu, M.}, \bibinfo{author}{Cao, J.}, \bibinfo{author}{Wang, X.}, \bibinfo{author}{Zhang, N.}, \bibinfo{year}{2021}.
\newblock \bibinfo{title}{Multi-attribute group decision-making considering opinion dynamics}.
\newblock \bibinfo{journal}{Expert Systems with Applications} \bibinfo{volume}{184}, \bibinfo{pages}{115479}.
\newblock \DOIprefix\doi{https://doi.org/10.1016/j.eswa.2021.115479}.
\bibitem[{Lipiecki and Sznajd-Weron(2022)}]{Lip:Szn:22}
\bibinfo{author}{Lipiecki, A.}, \bibinfo{author}{Sznajd-Weron, K.}, \bibinfo{year}{2022}.
\newblock \bibinfo{title}{Polarization in the three-state q-voter model with anticonformity and bounded confidence}.
\newblock \bibinfo{journal}{Chaos, Solitons \& Fractals} \bibinfo{volume}{165}, \bibinfo{pages}{112809}.
\newblock \DOIprefix\doi{https://doi.org/10.1016/j.chaos.2022.112809}.
\bibitem[{Liu et~al.(2023)Liu, Chai and Li}]{Qin:Li:Min:23}
\bibinfo{author}{Liu, Q.}, \bibinfo{author}{Chai, L.}, \bibinfo{author}{Li, M.}, \bibinfo{year}{2023}.
\newblock \bibinfo{title}{Dynamics of expressed and private opinion evolution over issue sequences}.
\newblock \bibinfo{journal}{IEEE Transactions on Computational Social Systems} \bibinfo{volume}{10}, \bibinfo{pages}{2860 – 2869}.
\newblock \DOIprefix\doi{10.1109/TCSS.2022.3222443}. \bibinfo{note}{cited by: 2}.
\bibitem[{Ma and Zhang(2021)}]{Ma:Zha:21}
\bibinfo{author}{Ma, S.}, \bibinfo{author}{Zhang, H.}, \bibinfo{year}{2021}.
\newblock \bibinfo{title}{Opinion expression dynamics in social media chat groups: An integrated quasi-experimental and agent-based model approach}.
\newblock \bibinfo{journal}{Complexity} \bibinfo{volume}{2021}, \bibinfo{pages}{1--14}.
\newblock \DOIprefix\doi{10.1155/2021/2304754}.
\bibitem[{MacDonald et~al.(2011)MacDonald, Nail and Harper}]{Mac:Nai:Har:11}
\bibinfo{author}{MacDonald, G.}, \bibinfo{author}{Nail, P.}, \bibinfo{author}{Harper, J.}, \bibinfo{year}{2011}.
\newblock \bibinfo{title}{Do people use reverse psychology? an exploration of strategic self-anticonformity}.
\newblock \bibinfo{journal}{Social Influence} \bibinfo{volume}{6}, \bibinfo{pages}{1--14}.
\newblock \DOIprefix\doi{10.1080/15534510.2010.517282}.
\bibitem[{Manfredi et~al.(2020)Manfredi, Guazzini, Roos, Postmes and Koudenburg}]{Man:etal:20}
\bibinfo{author}{Manfredi, R.}, \bibinfo{author}{Guazzini, A.}, \bibinfo{author}{Roos, C.A.}, \bibinfo{author}{Postmes, T.}, \bibinfo{author}{Koudenburg, N.}, \bibinfo{year}{2020}.
\newblock \bibinfo{title}{Private-public opinion discrepancy}.
\newblock \bibinfo{journal}{{PLOS} {ONE}} \bibinfo{volume}{15}, \bibinfo{pages}{e0242148}.
\newblock \DOIprefix\doi{10.1371/journal.pone.0242148}.
\bibitem[{Martins(2008)}]{Mar:08}
\bibinfo{author}{Martins, A.C.R.}, \bibinfo{year}{2008}.
\newblock \bibinfo{title}{Continuous opinions and discrete actions in opinion dynamics problems}.
\newblock \bibinfo{journal}{International Journal of Modern Physics C} \bibinfo{volume}{19}, \bibinfo{pages}{617--624}.
\newblock \DOIprefix\doi{10.1142/S0129183108012339}.
\bibitem[{Nail et~al.(2013)Nail, Domenico and MacDonald}]{Nai:Dom:Mac:13}
\bibinfo{author}{Nail, P.R.}, \bibinfo{author}{Domenico, S.I.D.}, \bibinfo{author}{MacDonald, G.}, \bibinfo{year}{2013}.
\newblock \bibinfo{title}{Proposal of a double diamond model of social response}.
\newblock \bibinfo{journal}{Review of General Psychology} \bibinfo{volume}{17}, \bibinfo{pages}{1--19}.
\newblock \DOIprefix\doi{10.1037/a0030997}, \href{http://arxiv.org/abs/https://doi.org/10.1037/a0030997}{\tt arXiv:https://doi.org/10.1037/a0030997}.
\bibitem[{Noorazar(2020)}]{Noo:20}
\bibinfo{author}{Noorazar, H.}, \bibinfo{year}{2020}.
\newblock \bibinfo{title}{Recent advances in opinion propagation dynamics: a 2020 survey}.
\newblock \bibinfo{journal}{The European Physical Journal Plus} \bibinfo{volume}{135}.
\newblock \DOIprefix\doi{10.1140/epjp/s13360-020-00541-2}.
\bibitem[{Roy and Biswas(2022)}]{Roy:Bis:21}
\bibinfo{author}{Roy, S.}, \bibinfo{author}{Biswas, S.}, \bibinfo{year}{2022}.
\newblock \bibinfo{title}{Opinion dynamics: public and private}.
\newblock \bibinfo{journal}{Philosophical Transactions of the Royal Society A: Mathematical, Physical and Engineering Sciences} \bibinfo{volume}{380}.
\newblock \DOIprefix\doi{10.1098/rsta.2021.0169}.
\bibitem[{Shang(2021)}]{Sha:21}
\bibinfo{author}{Shang, Y.}, \bibinfo{year}{2021}.
\newblock \bibinfo{title}{Resilient consensus for expressed and private opinions}.
\newblock \bibinfo{journal}{IEEE Transactions on Cybernetics} \bibinfo{volume}{51}, \bibinfo{pages}{318--331}.
\newblock \DOIprefix\doi{10.1109/TCYB.2019.2939929}.
\bibitem[{Sznajd-Weron et~al.(2024)Sznajd-Weron, Jędrzejewski and Kamińska}]{Szn:Jed:Kam:24}
\bibinfo{author}{Sznajd-Weron, K.}, \bibinfo{author}{Jędrzejewski, A.}, \bibinfo{author}{Kamińska, B.}, \bibinfo{year}{2024}.
\newblock \bibinfo{title}{Toward understanding of the social hysteresis: Insights from agent-based modeling}.
\newblock \bibinfo{journal}{Perspectives on Psychological Science} \bibinfo{volume}{19}, \bibinfo{pages}{511--521}.
\newblock \DOIprefix\doi{10.1177/17456916231195361}.
\bibitem[{Wang et~al.(2020)Wang, He, Huang and Ma}]{Wan:etal:20}
\bibinfo{author}{Wang, W.T.}, \bibinfo{author}{He, Y.L.}, \bibinfo{author}{Huang, J.Z.}, \bibinfo{author}{Ma, L.H.}, \bibinfo{year}{2020}.
\newblock \bibinfo{title}{A new approach to solve opinion dynamics on complex networks}.
\newblock \bibinfo{journal}{Expert Systems with Applications} \bibinfo{volume}{145}, \bibinfo{pages}{113132}.
\newblock \DOIprefix\doi{10.1016/j.eswa.2019.113132}.
\bibitem[{Xia et~al.(2023)Xia, Liang and Ye}]{Xia:Ho:Men:23}
\bibinfo{author}{Xia, W.}, \bibinfo{author}{Liang, H.}, \bibinfo{author}{Ye, M.}, \bibinfo{year}{2023}.
\newblock \bibinfo{title}{Asynchronous expressed and private opinion dynamics on influence networks}.
\newblock \bibinfo{journal}{IEEE Transactions on Control of Network Systems} \bibinfo{volume}{10}, \bibinfo{pages}{544 – 555}.
\newblock \DOIprefix\doi{10.1109/TCNS.2022.3219766}. \bibinfo{note}{cited by: 0}.
\bibitem[{Ye et~al.(2019)Ye, Qin, Govaert, Anderson and Cao}]{Ye:etal:19}
\bibinfo{author}{Ye, M.}, \bibinfo{author}{Qin, Y.}, \bibinfo{author}{Govaert, A.}, \bibinfo{author}{Anderson, B.D.}, \bibinfo{author}{Cao, M.}, \bibinfo{year}{2019}.
\newblock \bibinfo{title}{An influence network model to study discrepancies in expressed and private opinions}.
\newblock \bibinfo{journal}{Automatica} \bibinfo{volume}{107}, \bibinfo{pages}{371--381}.
\newblock \DOIprefix\doi{10.1016/j.automatica.2019.05.059}.
\bibitem[{Zhang et~al.(2022)Zhang, Chen, Gao, Dong and Pedryczc}]{Zha:etal:22}
\bibinfo{author}{Zhang, Y.}, \bibinfo{author}{Chen, X.}, \bibinfo{author}{Gao, L.}, \bibinfo{author}{Dong, Y.}, \bibinfo{author}{Pedryczc, W.}, \bibinfo{year}{2022}.
\newblock \bibinfo{title}{Consensus reaching with trust evolution in social network group decision making}.
\newblock \bibinfo{journal}{Expert Systems with Applications} \bibinfo{volume}{188}, \bibinfo{pages}{116022}.
\newblock \DOIprefix\doi{https://doi.org/10.1016/j.eswa.2021.116022}.

\end{thebibliography}

\end{document}